\documentclass[aps,preprint,amsmath,amssymb,superscriptaddress,showpacs]{revtex4-1}
\usepackage{amsmath,amssymb,bm,graphicx,epsfig,psfrag,ulem,color,natbib}

\newcommand{\vv}{{\bf {v}}}
\newcommand{\rr}{{\bf {r}}}
\newcommand{\pp}{{\bf {p}}}

\newcommand{\sss}{{\bf {s}}}

\newcommand{\ep}{{\mbox{$\epsilon_p$}}}
\newcommand{\eF}{{\mbox{$\epsilon_F$}}}
\newcommand{\vF}{{\mbox{$v_F$}}}
\newcommand{\pF}{{\mbox{$p_F$}}}
\newcommand{\vvs}{{\bf {v}}_s}
\newcommand{\vg}{{\mbox{$v_g$}}}

\begin{document}
%\draft
%\def\pd#1#2{\frac{\partial #1}{\partial #2}}

\title{\bf Cross-sections of Andreev scattering by quantized vortex rings in $^3$He-B.}

\author{N.~Suramlishvili}
\author{A.~W.~Baggaley}
\author{C.~F.~Barenghi}\affiliation{School of Mathematics and Statistics, University of
Newcastle, Newcastle upon Tyne, NE1 7RU, UK}

\author{Y.A.~Sergeev}\affiliation{School of Mechanical and Systems Engineering, Newcastle
University, Newcastle upon Tyne, NE1 7RU, UK}
\date {\today}

\begin {abstract}
We studied numerically the Andreev scattering cross-sections of three-dimensional isolated
quantized vortex rings in superfluid $^3$He-B at ultra-low temperatures. 
We calculated the dependence of the cross-section on the ring's size and on 
the angle between the beam of incident thermal quasiparticle excitations and the 
direction of the ring's motion. We also introduced, and investigated numerically, 
the cross-section averaged over all possible orientations of the vortex ring; 
such a cross-section may be particularly relevant for the analysis of experimental data. 
We also analyzed the r\^ole of screening effects for Andreev reflection of quasiparticles
by systems of vortex rings. Using the results obtained for isolated rings
we found that the screening factor for a system of unlinked rings depends strongly on the average 
radius of the vortex ring, and that the
screening effects increase with decreasing the rings' size.
\end{abstract}

\maketitle

\section{Introduction}
\label{sec:intro}

Superfluid turbulence consists of a disordered tangle of quantized 
vortex filaments which move under the velocity
field of each other~\cite{Donnelly-book,BDV-book}. If the temperature, $T$ is
sufficiently smaller than the critical temperature, $T_c$, then the
normal fluid can be neglected and the vortices do not experience any
friction effects~\cite{BDV}. The simplicity of the vortex structures
(discrete vortex lines) makes superfluid turbulence a
remarkable fluid system, particularly when compared to turbulence in
ordinary fluids. Unfortunately, in contrast with ordinary turbulence, 
only few experimental techniques of flow visualization and detection of turbulent structures are 
available in superfluids at very low temperatures.
Superfluid turbulence experiments are currently performed in both
$^4$He \cite{Golov,Roche-Barenghi,McClintock,Skrbek,Roche} and in $^3$He-B
\cite{Fisher,Bradley2,Bradley3,Finne,Yano,Hosio}. The methods used in these two
liquids are different. In superfluid $^3$He-B, at temperatures $T\ll T_c$, a powerful experimental technique, based on the Andreev scattering of thermal quasiparticle excitations, can be used to detect the vortex filaments, see for example the review article~\cite{Fisher_review}. This technique, having been pioneered and developed at Lancaster
University~\cite{Fisher,Bradley2,Bradley3}, is now also used at Aalto University in Helsinki \cite{Hosio} for measurements of vortex configurations. The Andreev scattering technique makes use of the fact that the energy dispersion curve, $E=E({\bf p})$ of
quasiparticle thermal excitations of momentum $\bf p$ is tied to the reference frame of the superfluid. From the Galilean invariance it follows that in this reference frame the dispersion curve tilts,
becoming
$E({\bf p})+{\bf p\cdot v}_s$ \cite{Fisher_review}, where $\vvs$ is the superfluid velocity. Thus, for thermal excitations whose energies are greater than the Fermi energy, $\eF$ (such excitations are known as quasiparticles), one side of the vortex filament presents
a potential barrier and they are reflected back almost exactly,
becoming quasiholes (excitations whose energy is smaller than $\eF$); the other side of the vortex lets the quasiparticles go through.
Quasiholes are reflected or transmitted in the opposite way. The vortex thus casts a symmetric ``Andreev''
shadow for the quasiparticles at one side and for the quasiholes at other side, and
by measuring the flux of excitations one detects the presence of the vortex.

In our earlier works we developed a theory of ballistic propagation of thermal excitations near a single, rectilinear vortex filament in $^3$He-B \cite{BSS1}, and studied interactions of thermal quasiparticles with simple, two-dimensional vortex configurations, such as clusters of vortex points \cite{BSS2} and a gas of point vortices and/or vortex-antivortex pairs \cite{BSS3}. In the latter two works we found and investigated the phenomenon of the so-called `partial screening' when the Andreev shadow of a system of vortices is no longer equal to the sum of shadows of individual vortices. However,
the results following from our two-dimensional models should be 
regarded as qualitative rather than quantitative; a quantitative comparison with experimental
observations and Andreev scattering data requires a fully three-dimensional study of vortex systems
and quasiparticles trajectories.

This work is concerned with the Andreev scattering of thermal excitations by individual vortex rings 
in three-dimensional geometry. Our study is particularly motivated 
by experimental observations of the transition from a gas of vortex rings to a 
dense vortex tangle \cite{Bradley2}, and measurements of
the decay of quantum turbulence generated by a vibrating grid shedding quantized vortex rings in alternating directions \cite{Bradley3}. Most conveniently the Andreev scattering of thermal excitation by quantized vortex rings can be characterized by the cross-section defined either by the ratio of the total number of quasiparticles reflected by the ring per unit time to the number flux density of quasiparticles incident on the ring, or, alternatively, by the cross-section defined as the ratio of the total power reflected by the ring to the flux density of energy carried by incident quasiparticles. In this work both cross-sections are calculated as functions of the ring's size and orientation with respect to the direction of the incoming beam of thermal excitations. Also calculated are the cross-sections averaged over all possible orientations of the ring.

The plan of the paper is the
following. In Sec.~\ref{sec:quasiparticles}, we shall introduce the equations of motion for ballistic quasiparticles in the superflow field, formulate the equations governing the fluid flow and the motion of quantized vortex rings, and define the cross-sections of interactions between thermal quasiparticles and vortex rings. In Sec.~\ref{sec:numerical} we describe
the numerical method. In Sec.~\ref{sec:crosssections} we shall calculate the scattering cross sections of the vortex rings and their systems.
In Sec.~\ref{sec:conclusions} we shall draw the conclusions.

%%%%%%%%%%%%%%%%%%%%%%%%%%%%%%%%%%%%%%%%%%%%%%%%%%%%%%%%%%%%%%%%%%%%%%%%%%%%%%%%%%%%%%%%%%%%%%%%%%%%%%%%%%%%%%%%%%%%%%%%%%%%%%%%%%

\section{Ballistic quasiparticles and cross-sections of the Andreev scattering in the flow field of quantized vortex ring}
\label{sec:quasiparticles}

We will be concerned with the
propagation of thermal excitations in $^3$He-B at temperatures $T\ll T_c$, where $T_c\approx1\,{\rm mK}$ is the critical temperature. Below all numerical data are taken at 0 bar pressure.

Neglecting spatial variations of
the order parameter, the energy of a thermal excitation of momentum $\pp$ in the flow field $\vvs(\rr,\,t)$ generated by the quantized vortex ring is
\begin{equation}
E(\pp,\,\rr,\,t)=\sqrt{\ep^2+\Delta_0^2}+\pp\cdot\vvs(\rr,\,t), 
\label{eq:energy}
\end{equation}
where
\begin{equation} 
\ep=\frac{p^2}{2m^*}-\eF
\label{eq:ep}
\end{equation}
is the ``kinetic'' energy of a thermal excitation
relative to the Fermi energy $\eF\approx2.27\times10^{-16}\,{\rm erg}$, $p=\vert\pp\vert$, $m^*\approx 3.01\times m=1.51 \times
10^{-23}~\rm g$ is the effective mass of excitation
in $^3$He-B (with $m$ being the bare mass of the $^3$He atom). We will be considering the propagation of thermal excitations at distances from the vortex core exceeding the zero-temperature 
coherence length, $\xi_0\approx0.75\times10^{-5}\,{\rm cm}$ so that the superfluid energy gap can be regarded as constant,
$\Delta_0=1.76k_BT_c\approx2.43\times 10^{-19}\rm erg$ (here $k_B$ is the Boltzmann's constant). Excitations with $\ep>0$ and $\ep<0$ are called, respectively, quasiparticles and quasiholes.

Below we will follow the approach developed in our earlier works~\cite{BSS1,BSS2,BSS3} and assume that the 
interaction term, $\pp\cdot\vvs$, varies on a spatial scale which is larger than 
$\xi_0=\hbar\vF/\pi\Delta_0$, where $\vF=\sqrt{2\eF/m^*}\approx5.48\times10^3\,{\rm cm/s}$ is the Fermi velocity. Then, following Refs.~\cite{Leggett,Yip}, Eq.~(\ref{eq:energy}) can be regarded as a semi-classical Hamiltonian for the excitation considered as a compact object (quasiparticle), whose 
position and momentum are $\rr(t)$ and $\pp(t)$ respectively, yielding the equations of motion
\begin{equation}
\dot{\rr}=\frac{\partial E}{\partial\pp}=\frac{\ep}{\sqrt{\ep^2+\Delta_0^2}}\frac{\pp}{m^*}+\vvs\,,
\label{eq:rmotion}
\end{equation}
\begin{equation}
\dot{\pp}=-\frac{\partial E}{\partial\rr}=-\frac{\partial}{\partial\rr}[\pp\cdot\vvs]\,,
\label{eq:pmotion}
\end{equation}
where a dot denotes a derivative with respect to time.
Note that the right-hand-side of Eq.~(\ref{eq:rmotion}) represents the group velocity of thermal quasiparticle.

In Eqs.~(\ref{eq:rmotion})-(\ref{eq:pmotion}), $\vvs$ represents the flow field generated by the quantized vortex ring. In the zero-temperature limit, the ring of radius $R$ moves in the direction orthogonal to the ring's plane with the self-induced velocity (here we assume that the vortex core is hollow)
\begin{equation}
v_i=\frac{\kappa}{2\pi R}\left[\ln\left(\frac{8R}{a_c}\right)-\frac{1}{2}\right]\,,
\label{eq:vi}
\end{equation}
where $\kappa=\pi\hbar/m=0.662\times10^{-3}\,{\rm cm^2/s}$ is the quantum of circulation in $^3$He-B, and $a_c$ is the core radius. Since $a_c$, being of the order of coherence length, $\xi_0$, is much smaller than the radius of the ring, it is appropriate to describe vortex lines as space curves of infinitesimal thickness.

The details of the fluid velocity field, $\vvs(\rr,\,t)$ generated by the vortex ring self-propagating in the inviscid fluid can be readily found for example in monograph by Lamb~\cite{Lamb}. However, for the purpose of this study it will be more convenient, using periodic boundary conditions, to calculate the flow field numerically from the
Biot-Savart law
\begin{equation}
\vvs(\rr,\,t)=-\frac{\kappa}{4\pi}\oint\frac{\rr-\sss}{\vert\rr-\sss\vert^3}\times d\sss\,,
\label{eq:biot_savart}
\end{equation}
where the integration extends over the whole vortex configuration. The motion and evolution of a single vortex ring or a system of quantized vortices is governed by the equation
\begin{equation}
\frac{d\sss}{dt}=\vvs(\sss,\,t)\,,
\label{eq:st}
\end{equation}
where $\sss=\sss(t)$ is a position of a point on the vortex line.

We consider the Andreev scattering of the net flux of excitations which results in the case where there is a (small) temperature gradient. Assuming that the 
source of thermal excitations is sufficiently far from quantized vortices, the beam of quasiparticles incident on the vortex ring (or the vortex tangle) can be regarded as one-dimensional. The differential fluxes of incident excitations, $\langle n\vg\rangle$ (${\rm cm^{-2}\, s^{-1}}$), and energy, $\langle n\vg E\rangle$ (${\rm erg\,cm^{-2}\, s^{-1}}$) (that is, respectively, the number of quasiparticles passing and the total energy carried by these quasiparticles through unit area) are \cite{Fisher_review,BSS1}
\begin{equation}
\langle n\vg\rangle=\int_{\Delta}^{\infty}N(E)\vg(E)\frac{\partial f(E)}{\partial T}\delta T\, dE\,,
\label{eq:ncurrent}
\end{equation}
\begin{equation}
\langle n\vg E\rangle=\int_{\Delta}^{\infty}N(E)\vg(E)E\frac{\partial f(E)}{\partial T}\delta T\, dE\,,
\label{eq:Ecurrent}
\end{equation}
where $\delta T\ll T$ is a temperature difference between the source of excitations and the opposite side of the experimental cell,
\begin{equation}
N(E)=N_F\frac{E}{(E^2-\Delta^2)^{1/2}}\,, \quad N_F=\frac{m\pF}{\pi^2\hbar^3}\,,
\label{eq:Ndensity}
\end{equation}
$N_F$ being the density of states at the Fermi energy with the corresponding Fermi momentum,
\begin{equation}
\vg=\frac{(E^2-\Delta^2)^{1/2}}{E}\vF
\label{eq:groupv}
\end{equation}
is the group velocity of Bogoliubov quasiparticle, and $f(E)$ is the Fermi distribution. At considered ultra-low temperatures, $T\leq0.15T_c$, typical of turbulence experiments in $^3$He-B, the Fermi distribution reduces to the Boltzmann distribution
\begin{equation}
f(E)=e^{-E/k_B T}\,.
\label{eq:Boltzmann}
\end{equation}

The quasiparticle trajectories resulting from interactions with the flow field of the vortex ring are determined from the solution of the problem represented by the closed system of equations (\ref{eq:rmotion}), (\ref{eq:pmotion}), (\ref{eq:biot_savart}), and (\ref{eq:st}) (the details of the numerical method will be discussed below in Sec.~\ref{sec:numerical}). The initial conditions follow from the Boltzmann distribution (\ref{eq:Boltzmann}) and the assumption that the initial positions of incident quasiparticles are distributed randomly on the plane orthogonal to the beam of excitations. The solution of this problem yields the total number of quasiparticles Andreev-reflected by the vortex configuration per unit time, $\dot{N}_R$ (${\rm s^{-1}}$), and the total power dissipated by Andreev-reflected quasiparticles, $Q_R$ (${\rm erg\,s^{-1}}$). Then, the cross-section of Andreev scattering by the vortex configuration (ring) can be defined as either
\begin{equation}
\sigma_N=\dot{N}_R/\langle n\vg\rangle
\label{eq:sigmaN}
\end{equation}
or
\begin{equation}
\sigma_E=Q_R/\langle n\vg E\rangle\,,
\label{eq:sigmaE}
\end{equation}
which we will call the particle and the thermal cross-section, respectively. Note that these cross-sections correspond to the area of Andreev shadow. Numerical calculations reported below in Sec.~\ref{sec:crosssections} show that $\sigma_N$ and $\sigma_E$ are practically indistinguishable in all considered situations, thus confirming that definitions (\ref{eq:sigmaN}) and (\ref{eq:sigmaE}) are correct.

\begin{figure*}
\centering
\begin{tabular}{ccc}
\epsfig{file=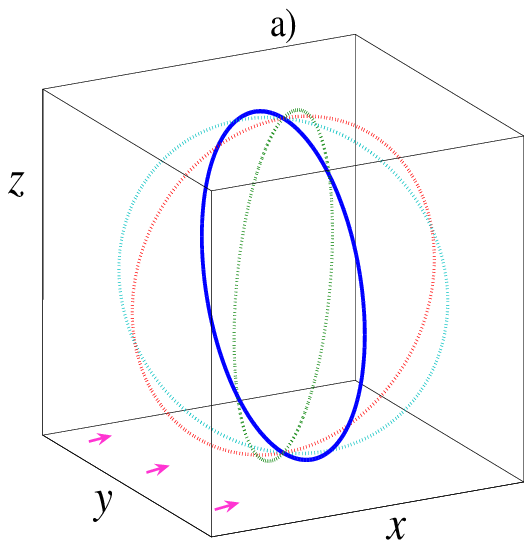,width=0.3\linewidth,clip=} &
\epsfig{file=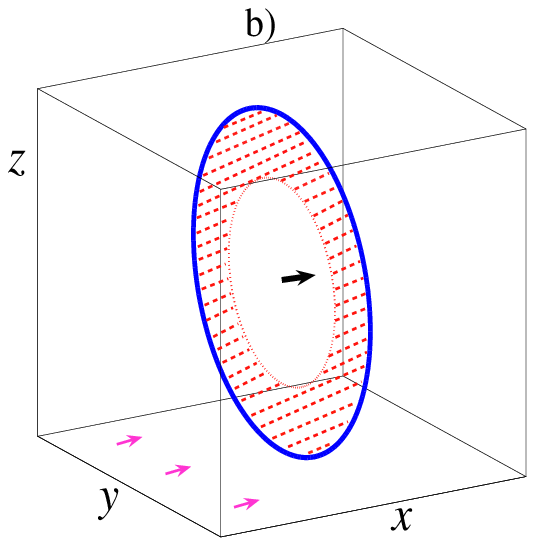,width=0.29\linewidth,clip=} &
\epsfig{file=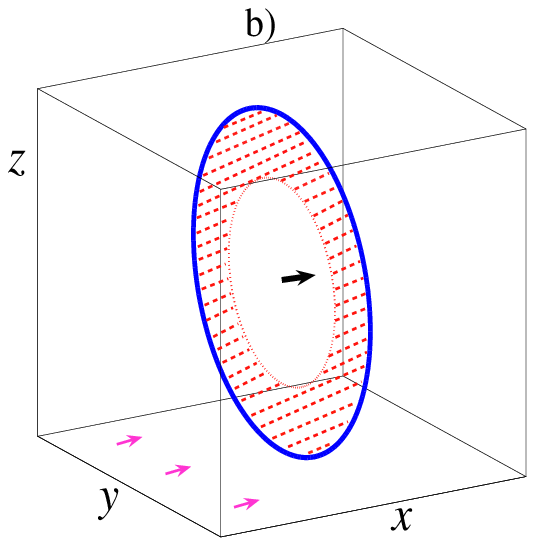,width=0.26\linewidth,clip=} 
\end{tabular}
\caption{(Color online) a) Orientations of the quantized vortex ring with respect to the 
x-direction (shown by three red arrows) of the beam of quasiparticles. Andreev shadow (red dashed area) of the ring (blue solid line) moving
b) parallel (in the positive x-direction, $\alpha=0$),
and c) antiparallel (in the negative x-direction, $\alpha=\pi$)
to the direction of the monochromatic beam of thermal excitations.}
\label{fig:ringshadows}
\end{figure*}

For the quantized vortex ring, the cross-section, $\sigma$ (below in this Section the subscript, $N$ or $E$ is omitted) is a function of the ring's size, $R$, the ring's velocity, which is itself a function of $R$, and the angle $\alpha$ between the beam of incident quasiparticles and the direction of translational motion of the ring, i.e. $\sigma=\sigma(R,\,\alpha)$. Different orientations of the ring with respect to the beam of excitations, and Andreev shadows in the cases where the ring moves either parallel or antiparallel to the direction of monochromatic beam of quasiparticles are illustrated on Fig.~\ref{fig:ringshadows}.

Of particular interest for interpretation of experiments will be the cross-section averaged over all possible angles $\alpha$. Assuming equal probability for all ring's orientations, the probability that the ring's velocity is at an angle between $\alpha$ and $\alpha+d\alpha$ with the direction of the beam can be easily calculated as $\frac{1}{2}\sin\alpha\,d\alpha$, with $0\leq\alpha\leq\pi$. Therefore, the angle-average cross-section of the ring of radius $R$ should be calculated as
\begin{equation}
\langle\sigma\rangle_{\alpha}=\frac{1}{2}\int_0^{\pi}\sigma(R,\,\alpha)\sin{\alpha}\,d\alpha\,.
\label{eq:angaver}
\end{equation}

\section{Numerical Method}
\label{sec:numerical} 

The superfluid velocity field, $\vvs(\rr,\,t)$ is calculated from Eqs.~(\ref{eq:biot_savart})-(\ref{eq:st}) by means of the vortex filament method using 
periodic boundary conditions. Calculations were performed in cubic periodic boxes of two sizes: $a=1.52\times10^{-2}$ and $a=2.25\times10^{-2}\,{\rm cm}$. The time evolution of the vortex configuration and, consequently, the velocity field are
calculated by the second-order Adams-Bashforth method using the fixed time step $\Delta t_v=2\times10^{-4}\,{\rm s}$. The technique of discretization of the vortex lines and the regularization of the Biot-Savart integral are standard. The details of our numerical algorithm are given in Ref.~\cite{Baggaley}.

Propagation of thermal excitations in the velocity field $\vvs(\rr,\,t)$ is governed by 
Eqs~(\ref{eq:rmotion})-(\ref{eq:pmotion}), which are solved using the numerical code based on 
the variable-step, variable-order implementation of 
the Numerical Differentiation Formulas (NDFs). A comprehensive and detailed description of the NDFs method is given in Ref.~\cite{Shampine}.

We tested our numerical method for Andreev reflection of quasiparticles, whose initial momentum 
is in the $x$-direction, by a single rectilinear vortex line located at
$x=y=0$ aligned along the z-direction; the velocity field of such a vortex is time independent. 
The energy of a thermal excitation,
defined by Eq.~(\ref{eq:energy}), and its $\hat z$-component of the orbital 
angular momentum, $J_z=p_xy-p_yx$ are both integrals of motion.
\begin{figure}
\centering \epsfig{figure=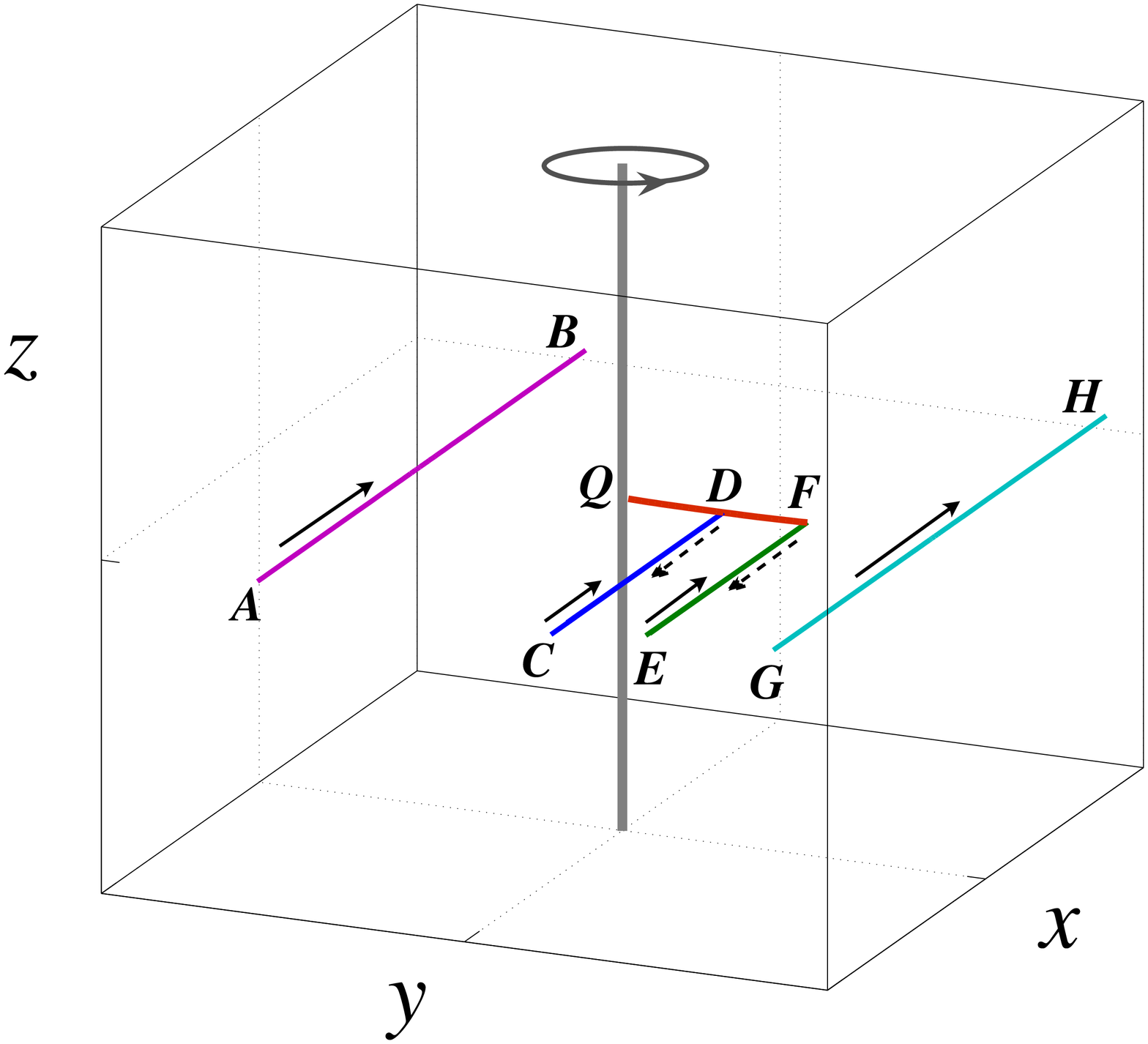,height=2.2in,angle=0}
\caption{(Color online)
The rectilinear line vortex (grey vertical line) and trajectories of the quasiparticles with average energy $\langle E\rangle=\Delta_0+k_BT$. The box shown in this figure is for visualization only -- it is not the
computational box; its sizes are $-5\times10^{-2}\leq x\leq5\times10^{-2}$, $-4\times10^{-3}\leq y\leq4\times10^{-3}$, and $-6\times10^{-2}\leq z\leq6\times10^{-2}$ (units in cm).
The purple (AB), blue (CD), green (EF), and navy (GH) lines are the excitation's trajectories corresponding to the initial impact parameters
$\rho_0\simeq 0.0029,-0.0011,~-0.002~{\rm and}-0.003\,\rm cm$ respectively. The grey ellipse indicates the direction of the velocity field
of the vortex, and the perpendicular to the vortex axis red (QF) line shows the extension ($\approx 0.002\,\rm cm$) of the shadow casted by
the vortex for the quasiparticles with energy $\langle E\rangle$. The solid and dashed arrows indicate the directions of motion of incident quasiparticles and retroreflected quasiholes, respectively.}
\label{fig:line}
\end{figure}
Fig.~\ref{fig:line} illustrates trajectories of excitations with different starting conditions identified by the impact parameter $\rho_0=y_0$, where $y_0$ is the initial $y$-coordinate of thermal excitation. (In other words, the impact parameter, $\rho_0$ is defined as a minimum distance between the rectilinear trajectory, which the ballistic quasiparticle would have followed in the absence of a vortex, and the vortex core, see Fig.~2 in Ref.~\cite{BSS1}.) Fig~\ref{fig:potential} shows 
that in this calculation the relative errors in the quasiparticle's energy, $\Delta E/E_0=(E-E_0)/E_0$ and momentum, $\Delta J/J_{z0}=(J_z-J_{z0})/J_{z0}$ (where $E_0$ and $J_{z0}$ are, respectively, 
the initial energy and $z$-component of momentum),
are less than $2.5\times 10^{-4}$ and $10^{-3}$, respectively, 
so that our method conserves the integrals of motion very well.
\begin{figure}
\centering
\begin{tabular}{cc}
\epsfig{file=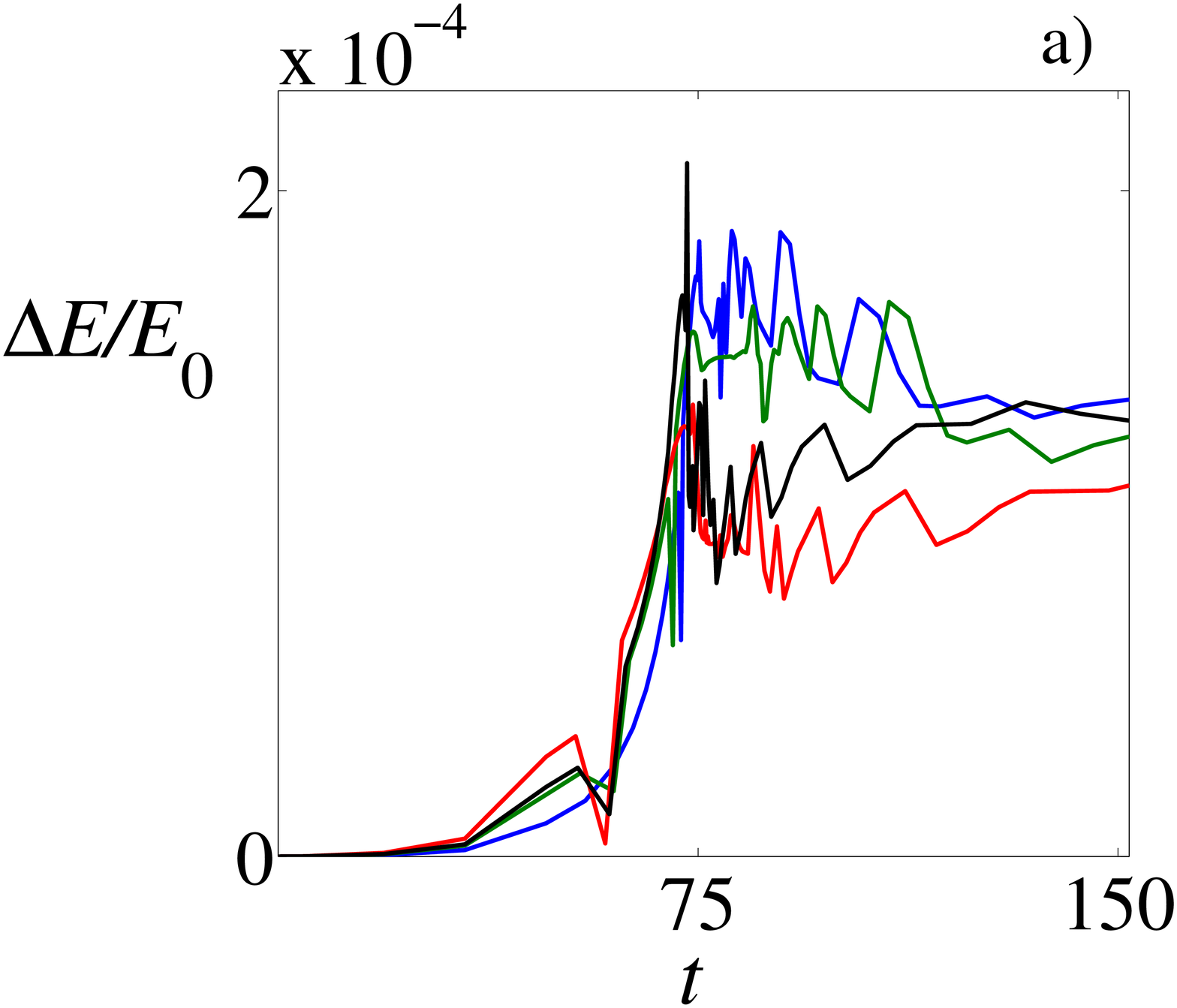,width=0.4\linewidth,clip=} &
\epsfig{file=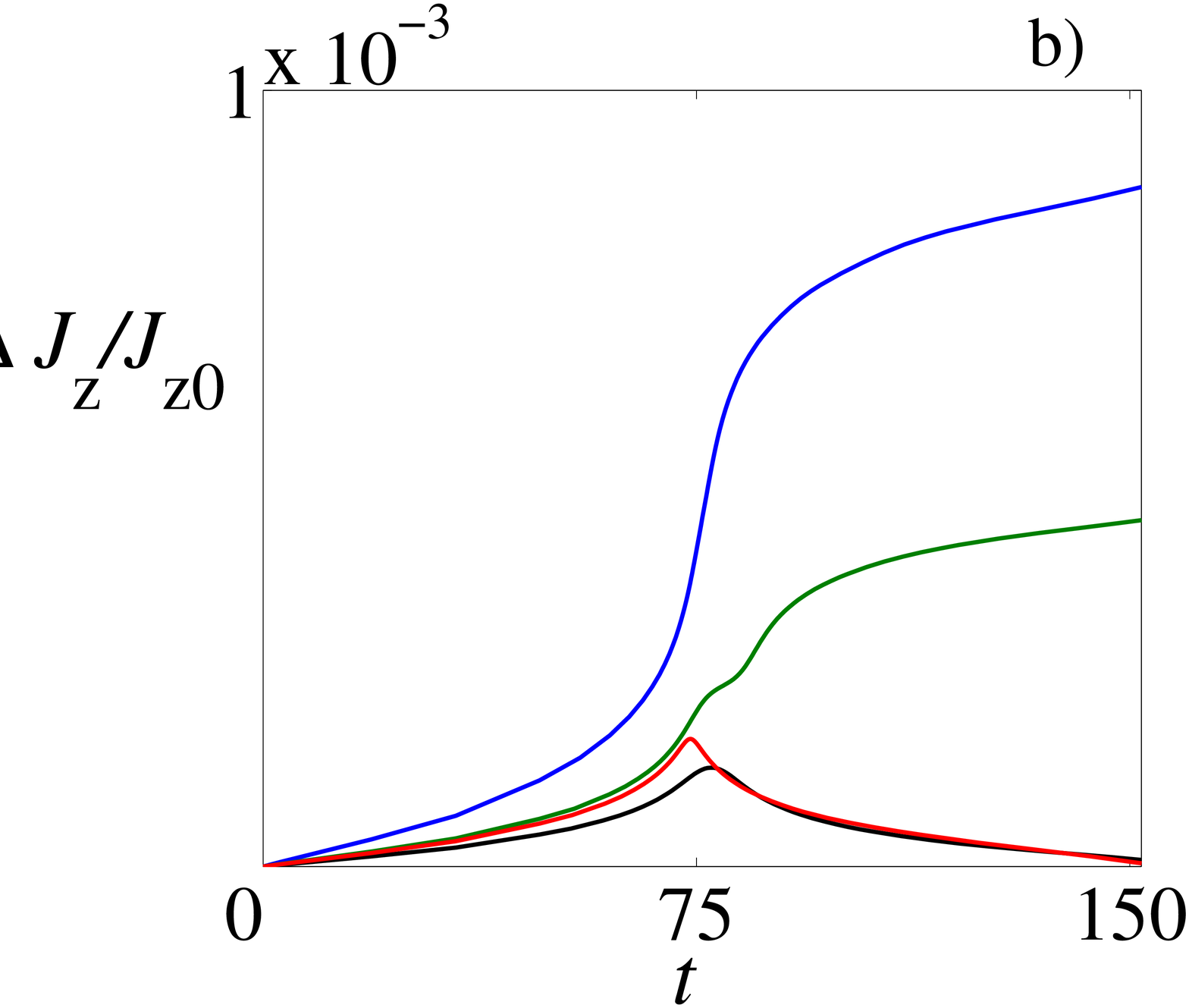,width=0.4\linewidth,clip=}
\end{tabular}
\caption{(Color online) a) Relative tolerance of the energy, $(E-E_0)/E_0$. b)~Relative tolerance of the $z$-component of orbital angular momentum,
$(J_z-J_{0z})/J_{0z}$. Time is in units of $\mu s$. The red, black, blue and green lines correspond 
to the trajectories $AB$, $GHG$, $CDC$ and $EF$ of Fig.~\ref{fig:line}, respectively.}
\label{fig:potential}
\end{figure}

To test our numerical method in the three-dimensional case for moving rings or their systems, we note first that in such a case the energy and the $z$-projection
of the orbital angular momentum are no longer integrals of motion.
However, they nearly are because the time scales of the quasiparticle
motion and of the fluid motion differ by at least an order of 
magnitude so that the flow field can be regarded as frozen, see the next paragraph. Our
calculations show that, in the three-dimensional case, for a single ring as well as 
for a system of several rings the relative errors in energy and in
the orbital angular momentum remain sufficiently small in all considered 
situations, $\Delta E/E_0<1.5\times10^{-2}$ and 
$\Delta J_z/J_{z0}<1.5\times10^{-2}$, respectively. These estimates have
been obtained for the smallest of considered rings, with $R=R_{min}=7.5\times10^{-5}\,{\rm cm}$.
Clearly, the errors becomes smaller with increasing the ring's
size.

In principle, Eqs.~(\ref{eq:biot_savart})-(\ref{eq:st}) and (\ref{eq:rmotion})-(\ref{eq:pmotion}) should be regarded as a system of equations whose solution yields simultaneously the time-dependent vortex configuration, the fluid velocity field, and the trajectories of quasiparticle thermal excitations. A numerical solution of such a system of equations presents formidable difficulties, in particular because the equations (\ref{eq:rmotion})-(\ref{eq:pmotion}) governing the motion of quasiparticles are stiff. 
However, because a typical time of travel of an excitation within the computational box 
is much smaller than the characteristic timescale of the motion and evolution of quantized vortices, the problem can be reduced to simpler calculations of quasiparticle motion in the frozen flow field of vortex configuration at any instant of time. To justify this approach we calculated, in the computational box of the larger size, $a=2.25\times10^{-2}\,{\rm cm}$, the time of travel of the excitation in the velocity field of the rectilinear vortex illustrated in Fig.~\ref{fig:line}. We found that the longest average time, which is of the order of $10^{-3}\,{\rm s}$, is spent within the box by quasiparticles
whose initial momentum is $p\approx\pF+2\times10^{-5}\pF$, where $\pF=\sqrt{2m^*\eF}\approx8.28\times10^{-2}\,{\rm g\,cm/s}$ is the Fermi momentum. The average group velocity of these, slowest quasiparticles is about $10^2\,{\rm cm/s}$. On the other hand, the largest velocity of the vortex points can be estimated 
as $v_\ell\sim\kappa/(2\pi R_{min})$, where $R_{min}$ is the radius of the smallest vortex ring. 
In our calculations $R_{min}=7.5\times10^{-5}\,{\rm cm}$, so that $v_{\ell}\sim10\,{\rm cm/s}$, 
which is an order of magnitude smaller than the average group velocity of the slowest quasiparticles. 
This justifies the `frozen flow field' approach for solving Eqs~(\ref{eq:rmotion})-(\ref{eq:pmotion}).

In our numerical simulations the flux of thermal excitations is modelled by $N_{qs}=52272$ quasiparticles entering from one side of the computational box and moving parallel 
to the $x$-direction. Initial positions on the $(y,z)$-plane and energies of quasiparticles, $E_0~(\Delta_0<E_0\leq 1.7\Delta_0)$ 
are uniformly distributed. The quasiparticles are characterized by the set of three integer numbers, $(n,\,m,\,k)$, where $n$ and $m$ refer
to the initial position of quasiparticle on the $(y,z)$-plane as follows:
\begin{equation}
y_n=-\frac{a}{2}+n\delta y_n\,,~~~~z_m=-\frac{a}{2}+m\delta z_m\,;~~~n,\,m=1,...,N,
\label{eq:qposition}
\end{equation}
where $N=66$, $a$ is the size of the cube, $\delta y_n$ and $\delta z_n$ are the distances, in $y$ and $z$ directions, respectively, between the nearest quasiparticles. The third number, $k$ refers to the energy level corresponding to the discrete momentum, $p_k=p_F+k\delta p$, (where $k=1,...,N_k$) and is
calculated as
\begin{equation}
E_{nmk}=\sqrt{(\ep)_k^2+\Delta_0^2}+\pp_k\cdot\vvs(-a/2,\,y_n,\,z_n)\,,
\label{eq:enerlevel}
\end{equation}
where $(\ep)_k=p_k^2/(2m^*)-\eF$. In our calculations $N_k=12$ (so that $N\times N\times N_k=N_{qs}$).

The incident number and energy differential fluxes of quasiparticles 
are now calculated numerically as follows:
\begin{equation}
\left(\begin{array}{c} \langle n\vg\rangle \\ \langle n\vg E\rangle \end{array}\right)
=\sum_{n,m,k}\frac{(\vg)_{nmk}}{\Delta_0}
\left(\begin{array}{c} E_{nmk} \\ E_{mk}^2 \end{array}\right)
e^{-17.6E_{nmk}/\Delta_0}\,,
\label{eq:fluxesnum}
\end{equation}
where, using the definition of the gap parameter, $\Delta_0$, for the considered temperature, $T=0.1T_c$ the exponent in the Boltzmann's factor has been replaced by $-17.6E/\Delta_0$, and the numerical approximation for the quasiparticle's group velocity, $\vv_g=\partial E/\partial\pp$ follows from Eqs.~(\ref{eq:rmotion}) and (\ref{eq:enerlevel}) in the form
\begin{equation}
(\vg)_{nmk}=\frac{(\ep)_k}{\sqrt{(\ep)_k^2+\Delta_0^2}}\frac{p_k}{m^*}
+v_s\left(-\frac{a}{2},\,y_n,\,z_m\right)\,.
\label{eq:vgroupnum}
\end{equation}
Having solved equations of motion (\ref{eq:rmotion})-(\ref{eq:pmotion}), the total number of quasiparticles Andreev-reflected per unit time, $\dot{N}_R$, and the total power dissipated by Andreev-reflected quasiparticles, $Q_R$, are calculated as
\begin{equation}
\left(\begin{array}{c} \dot{N}_R \\ Q_R \end{array}\right)
=\sum_{n,m,k}R_{nmk}\frac{(\vg)_{nmk}}{\Delta_0}
\left(\begin{array}{c} E_{nmk} \\ E_{mk}^2 \end{array}\right)
e^{-17.6E_{nmk}/\Delta_0}\,\Delta S_{nm}\,,
\label{eq:NQrefl}
\end{equation}
where $R_{nmk}=1$ if the $(n,\,m,\,k)$-th particle is Andreev-reflected, otherwise $R_{nmk}=0$. In Eq.~(\ref{eq:NQrefl}),
$\Delta S_{nm}=\frac{1}{4}(\delta y_n+\delta y_{n+1})(\delta z_m+\delta z_{m+1})$ is the area element of the $(y,z)$-plane.

As was mentioned earlier in this Section, the calculations of cross-sections
have been performed in computational boxes of two different sizes:
$a=1.52\times 10^{-2}\rm cm$ and $a=2.25\times 10^{-2}\rm cm$. Reported in the next Section~\ref{sec:crosssections}, the results of calculations of scattering cross-sections turn out to be independent of the size of computational box. This justifies both the correctness of definitions~(\ref{eq:sigmaN})-(\ref{eq:sigmaE}) and the accuracy of numerical approximations described in this Section.

%%%%%%%%%%%%%%%%%%%%%%%%%%%%%%%%%%%%%%%%%%%%%%%%%%%%%%%%%%%%%%%%%%%%%%%%%%%%%%%%%%%%%%%%%%%%%%%%%%%%%%%%%%%%%%%%%%%%%%%%%%%%%%%%%%%%%
\section{Results and discussions}
\label{sec:crosssections}

We start with the
calculation of the scattering cross section as a function of the radius of the quantized vortex ring, $R$, 
and the angle, $\alpha$ between the beam of incident quasiparticles and the direction of 
translational motion 
of the ring. The angle is $\alpha=0$ in the case where quasiparticles and the ring 
move in the same direction, and $\alpha=\pi$ if they move in opposite directions.

\begin{figure}
\centering \epsfig{figure=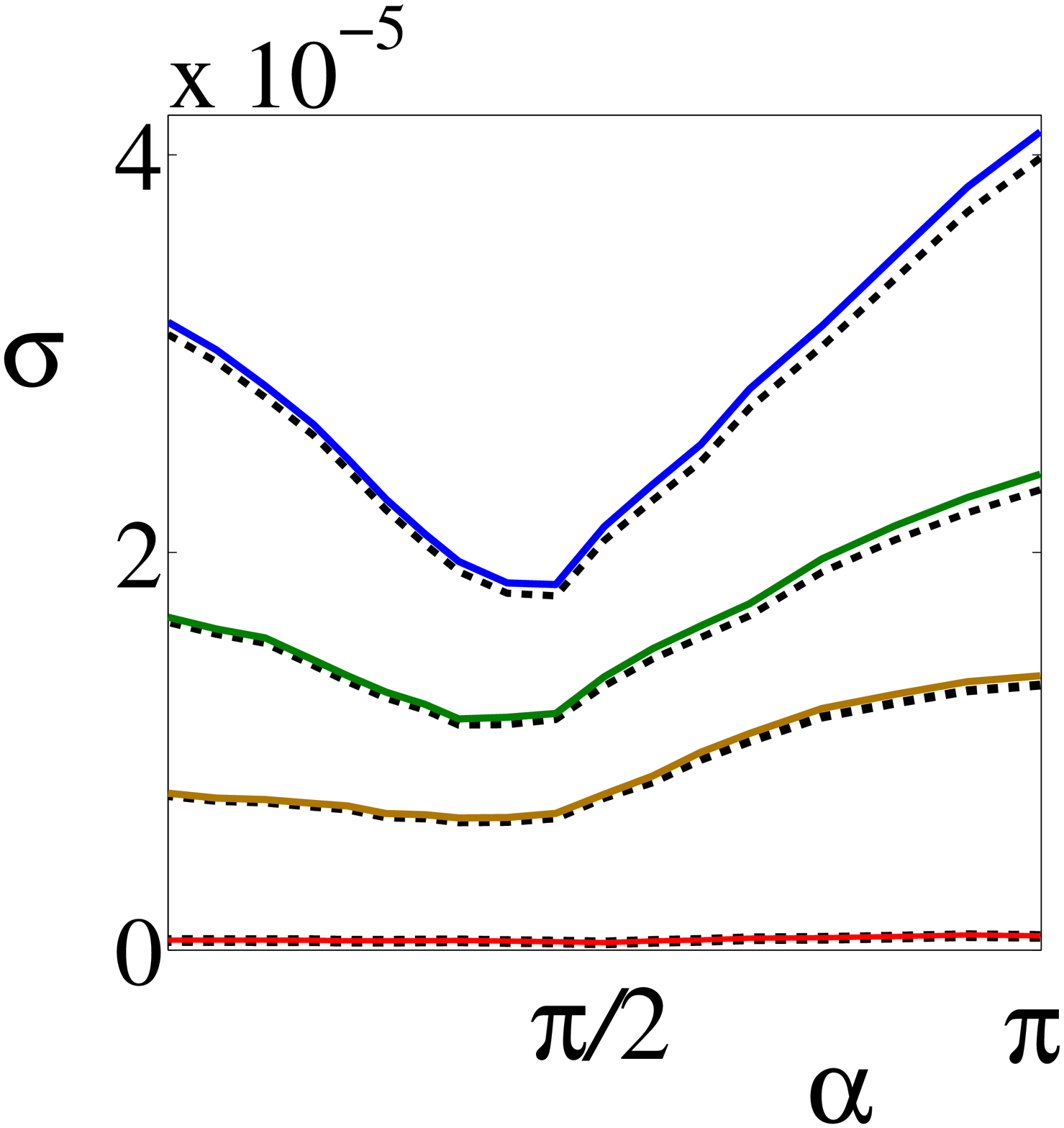,height=2in,angle=0}
\caption{(Color online)
 The cross-sections (in units of $\rm cm^2$) of Andreev scattering by an isolated vortex ring, $\sigma_N$ (solid lines) and $\sigma_E$ (dashed lines) as functions of the angle, $\alpha$ between the incident beam of excitations and the direction of motion of the vortex ring. The pairs of curves, from top to bottom, correspond to the ring's radii $R=3.9\times10^{-3}$, $2.4\times10^{-3}$, $1.4\times10^{-3}$, and $1.2\times10^{-4}\,{\rm cm}$.}
\label{fig:angledepend}
\end{figure}

The results of the calculation are shown in Fig.~\ref{fig:angledepend}.
As it has already been mentioned 
in Sec.~\ref{sec:quasiparticles}, the cross-sections $\sigma_N$ and $\sigma_E$, 
defined by formulae (\ref{eq:sigmaN}) and (\ref{eq:sigmaE}), respectively, practically coincide. 
This is hardly surprising considering that both of them correspond to the area where quasiparticles 
are Andreev reflected, i.e. to the area of Andreev shadow. Unless specified otherwise, 
in the remainder of this Section we will not distinguish between $\sigma_N$ and $\sigma_E$, 
hence omitting the subscripts ``$N$'' and ``$E$''. 
From the results shown in Fig.~\ref{fig:angledepend} it is seen that the cross-sections 
(and hence the Andreev reflection area) is the largest in the case where the ring moves exactly 
towards the source of excitations ($\alpha=\pi$); in the case where the direction of the beam 
and that of the ring's motion coincide ($\alpha=0$), the cross-section is slightly smaller. 
The minimum reflection area occurs for angles slightly smaller than $\alpha=\pi/2$. 
For small rings the cross-section is almost angle-independent; most likely, this is because 
at small intervortex distances the process of Andreev reflection becomes dominated by the 
partial screening effects investigated in our earlier works \cite{BSS2,BSS3}.

\begin{figure}
\centering \epsfig{figure=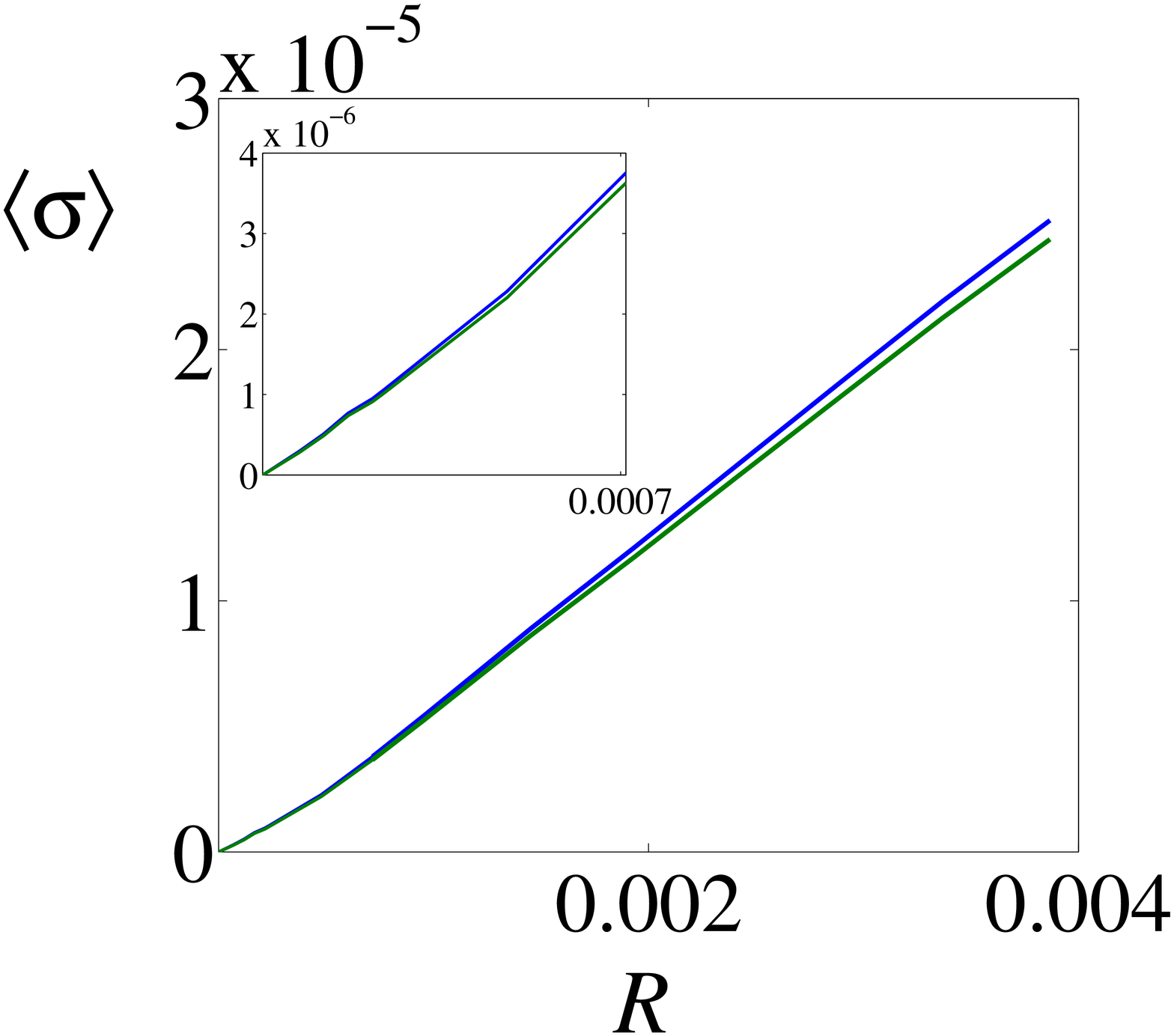,height=2in,angle=0}
\caption{(Color online) Angle-averaged particle (upper blue line) and thermal (lower green line) cross-sections ($\rm cm^2$) as functions of the radius of the ring (cm). 
Inset shows the behavior of $\sigma$ with $R$ for small rings.}
\label{fig:aversigmas}
\end{figure}

Fig.~\ref{fig:aversigmas} shows the dependence of the angle-average cross-section,
defined in Sec.~\ref{sec:quasiparticles} by Eq.~(\ref{eq:angaver}), on the radius of the ring, $R$. For sufficiently large rings, $R\gtrsim2.42\times10^{-4}\,{\rm cm}$ the angle-average cross-section, $\langle\sigma\rangle$ exhibits almost linear dependence on $R$, which can be approximated as
\begin{equation}
\langle\sigma\rangle\approx KR+C,
\label{eq:lindepend}
\end{equation}
with $K=6.4\times10^{-3}\,{\rm cm}$ and $C=-5.13\times10^{-7}\,{\rm cm^2}$. For smaller rings,
the behavior of $\langle\sigma\rangle$ with $R$ is also practically linear, with $K=4.3\times10^{-3}\,{\rm cm}$ and $C=0$. A rather sharp change of behavior occurring at $R\approx2.42\times10^{-4}\,{\rm cm}$ can again be attributed to significant contribution of partial screening effects for small rings~\cite{sizes}.

\begin{figure}
\centering \epsfig{figure=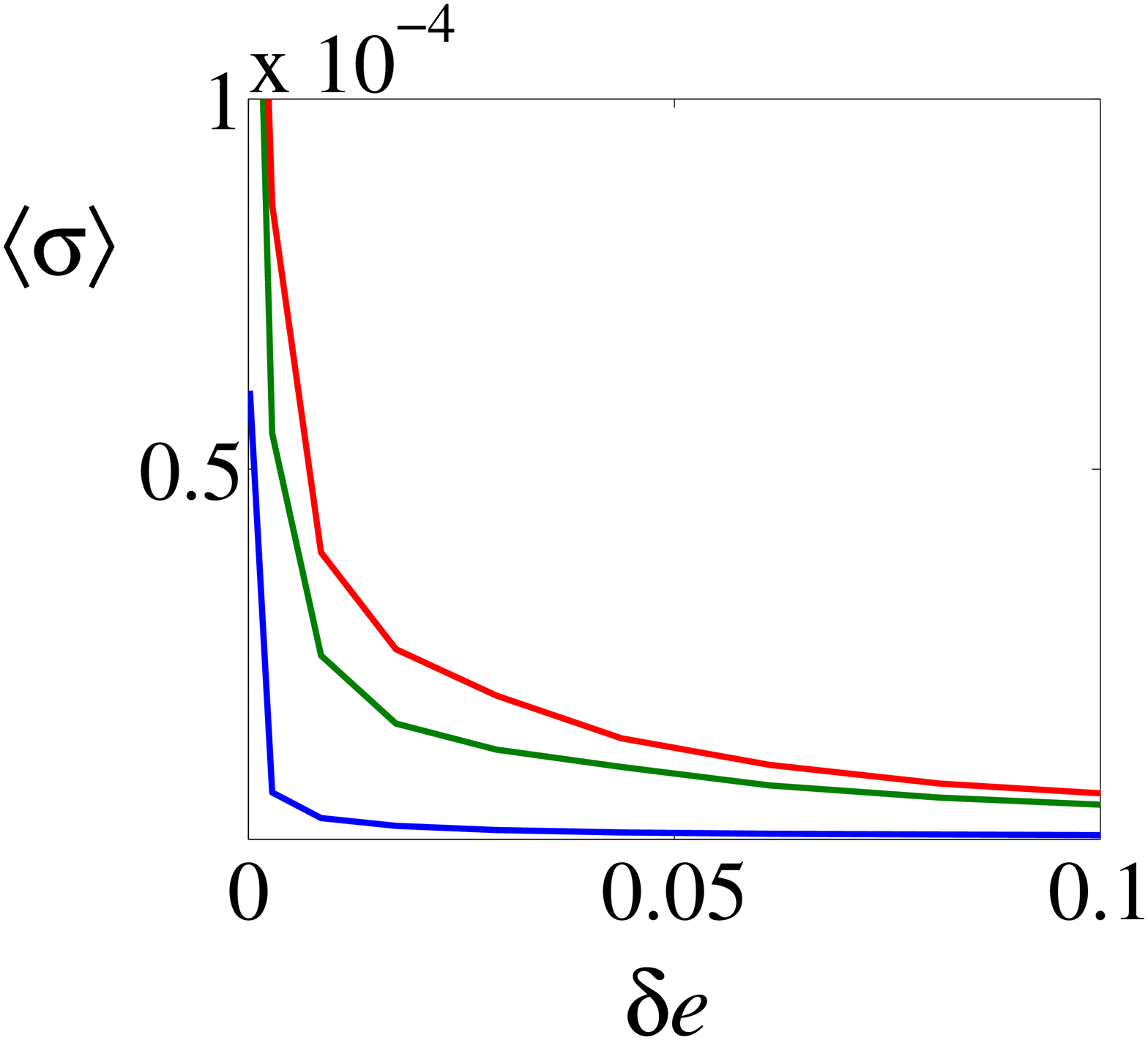,height=2.1in,angle=0}
\caption{(Color online) Scattering cross section ($\rm cm^2$) of a monochromatic beam of quasiparticles with energy $E$ vs dimensionless parameter $\delta e=(E-\Delta_0)/\Delta_0$
for radii of the ring, from top to bottom, $R=3.34\times 10^{-3}$, $R=2.41\times 10^{-3}$, and 
$R=4.83\times 10^{-4}\,\rm cm$.}
\label{fig:enersigmas}
\end{figure}

So far we analyzed the Andreev scattering of quasiparticles whose initial energies are uniformly distributed in the incident beam. Of a certain interest are also cross-sections of a monochromatic beam, i.e. such that all incident quasiparticles have the same fixed energy, $E$. Illustrated by Fig.~\ref{fig:enersigmas},
our numerical calculation shows a strong decrease of the angle-average cross-section with increasing the non-dimensional parameter $\delta e=(E-\Delta_0)/\Delta_0$ and decreasing the ring radius, $R$.

\begin{figure*}
\centering \epsfig{figure=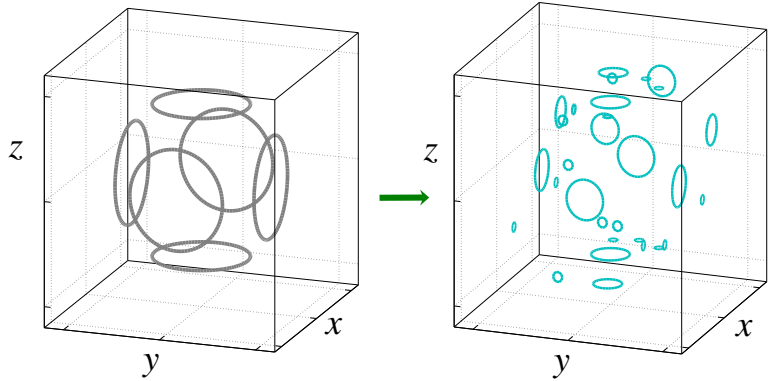,height=2in,angle=0}
\caption{(Color online) Modeling sequence of configurations of $n$ unlinked vorex rings with the total line length being preserved for all $n$. Initial configuration of six rings (left), and final configurations of 36
rings (right).}
\label{fig:loopsplit}
\end{figure*}

\begin{figure}
\centering
\begin{tabular}{cc}
\epsfig{file=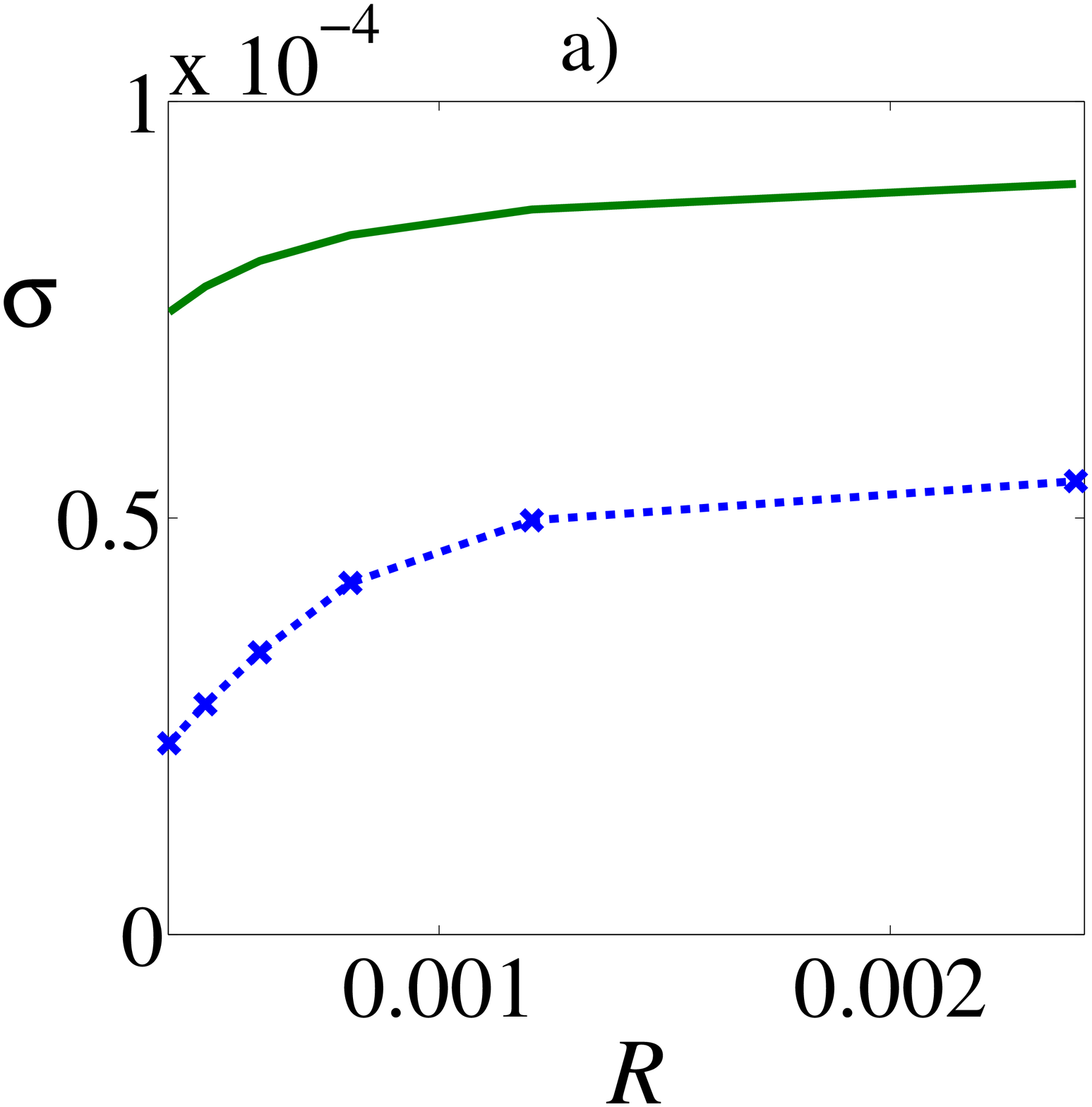,width=0.3\linewidth,clip=} &
\epsfig{file=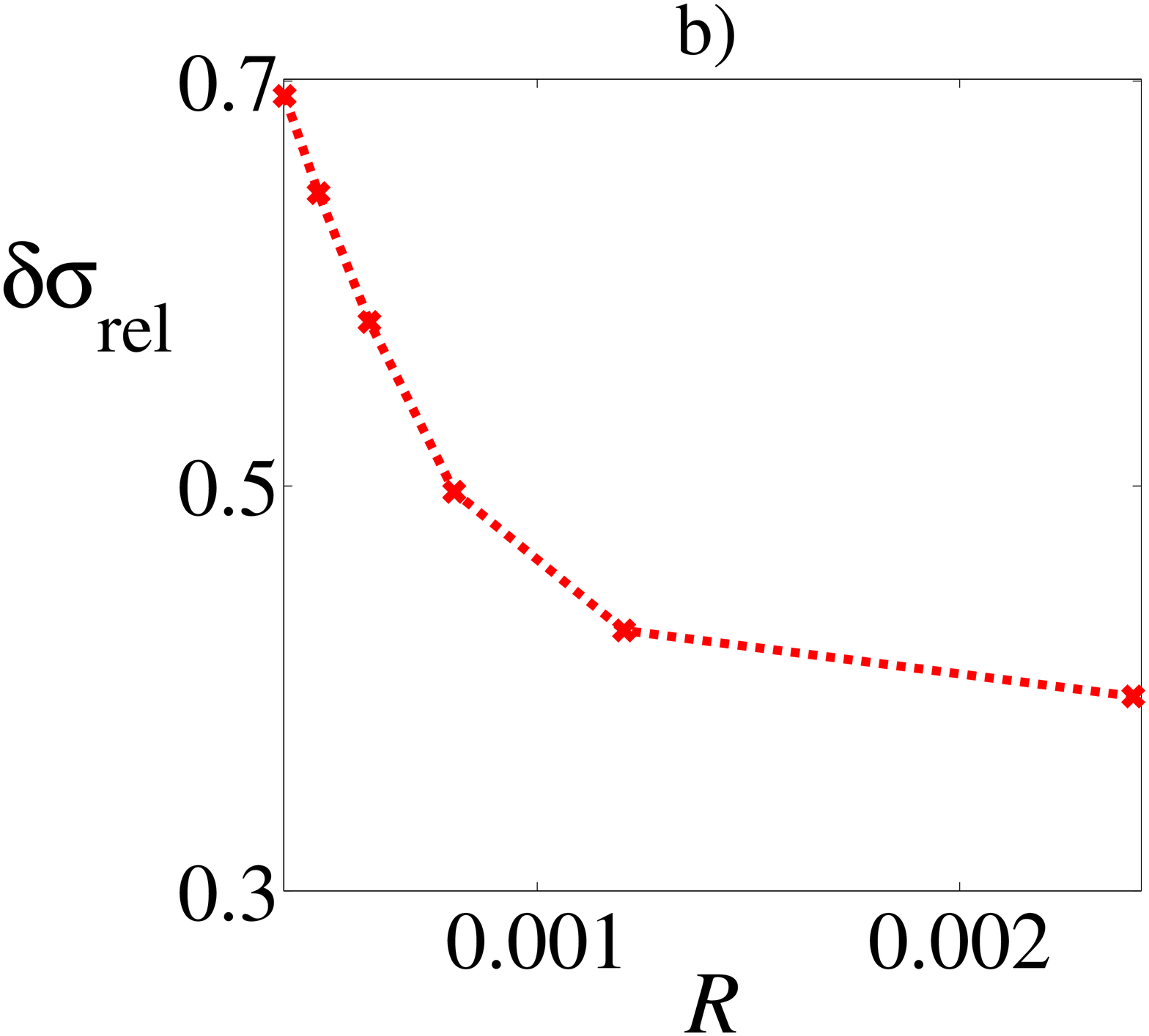,width=0.33\linewidth,clip=}
\end{tabular}
\caption{(Color online) Sum of the angle-average cross-sections of $n$ individual vortex rings, $n\langle\sigma(R)\rangle$ (upper green line, left), total (screened) cross-section, $\sigma_n$ (lower blue line, left), and the screening factor (right) as functions of radius (cm). For each $n$ all rings have the same radius, $R=L_{tot}/(2\pi n)$, with $L_{tot}$ being the same for all $n$. Units of cross-sections are in ${\rm cm^2}$.}
\label{fig:screen}
\end{figure}

Until now we have considered the Andreev reflection of quasiparticles on a single quantized vortex ring. Here we will analyze briefly the case where the incident beam of quasiparticles is Andreev-reflected by 
a system of $n$ unlinked quantized vortex rings. Such a system can also be characterized by the average radius of the ring, $\bar{R}=\sum R_i/n$. Enforcing the same total line length, $L_{tot}$ for all $n$, we calculate the cross-section of Andreev scattering by the system of vortex rings for $n$ progressively increasing from $n_i$ corresponding to a system of just a few large rings, to $n=n_f$ corresponding to a system of many smaller rings, see Fig.~\ref{fig:loopsplit}. In our calculation $n_i=6$, $n_f=36$, and $L_{tot}=0.908\times10^{-1}\,{\rm cm}$. 
We have to emphasize that such a sequence of configurations of vortex rings is not due to the Biot-Savart evolution of the system, but is the result of numerically enforced algorithm.
Our calculation shows that, due to the screening effects, the total cross-section of $n$ rings, $\sigma_n(\bar{R})$ is smaller than the sum of angle-averaged cross-sections of individual rings, that is
\begin{equation}
\sigma_n(\bar{R})<\sum_{i=1}^n\langle\sigma(R_i)\rangle\,.
\label{eq:total-aver-sigma}
\end{equation}
Numerical calculations also show that the screening factor, defined as
\begin{equation}
\delta\sigma_{rel}=1-\sigma_n(\bar{R})\left(\sum_{i=1}^n\langle\sigma(R_i)\rangle\right)^{-1}\,,
\label{eq:screening}
\end{equation}
increases with decreasing the average radius, $\bar{R}$. For each $n$, from $n_i=6$ to $n_f=36$, we analyzed a number of configurations of the system of vortex rings, and found that for each $n$ the total cross-section of the system oscillates around the value corresponding to the case where all rings have the same radius, $R=\bar{R}$. This case is investigated in some more detail, and the calculated values of the total cross-section, $\sigma_n$, the sum of angle-averaged 
cross-sections of individual vortex rings, $\sum\langle\sigma(R_i)\rangle=n\langle\sigma(R)\rangle$, and the screening factor, $\delta\sigma_{rel}=1-\sigma_n(R)/(n\langle\sigma(R)\rangle)$ are represented in Fig.~\ref{fig:screen} as functions of radius, $R=L_{tot}/(2\pi n)$.
Results shown in Fig.~\ref{fig:screen} (left) indicate that, despite the total vortex line length remains the same for all values of $R$, the total scattering cross-section of the system decreases substantially with radius. Figure~\ref{fig:screen} (right) shows
the dramatic increase of the screening from $41\%$ up to $69\%$ with decreasing the rings radii from $R=2.41\times 10^{-3}\,\rm cm$ to $R=0.402\times 10^{-3}\,\rm cm$.
This may be explained with the help of results illustrated by Fig.~\ref{fig:enersigmas} which indicate that the main contribution to the cross-sections of smaller rings is made by
the low energy quasiparticles. For the high energy quasiparticles sufficiently small rings are almost transparent. When the number of rings is increased so that the rings' sizes are reduced,
most of the low energy quasiparticles are reflected by the front-line rings, and just a small fraction of excitations reaches the rings in the bulk of the system; hence, because
most of the high energy quasiparticles are not Andreev reflected at all, the screening effect increases. 
 
%%%%%%%%%%%%%%%%%%%%%%%%%%%%%%%%%%%%%%%%%%%%%%%%%%%%%%%%%%%%%%%%%%%%%%%%%%%%%%%%%%%%%%%%%%%%%%%%%%%%%%%%%%%%%%%%%%%%%%%%%%%%
\section{Conclusions}
\label{sec:conclusions}

In conclusion, we have analyzed, for the first time, the three-dimensional Andreev reflection of thermal quasiparticle excitations by quantized vortex rings in $^3$He-B. The particle and thermal cross-sections (i.e. the Andreev reflection areas) of quantized vortex rings are defined and calculated; the results show a strong dependence of the cross-section on the angle between the incident beam of quasiparticles and the direction of motion of the vortex ring. It is also shown that the particle and the thermal cross-sections practically coincide. Of a primary interest for interpretation of experimental data is the cross-section averaged over all possible orientations of the vortex ring. This is calculated and its dependence on the size of vortex ring is analyzed in detail. It is apparent that the phenomenon of partial screening investigated in the authors' earlier works in two dimensions, plays a major r\^ole for rings of sufficiently 
small size in three dimensions. The results are generalized for the case of Andreev reflection by the system of vortex rings. It is found that due to the screening effects the total cross-section of the system of vortex rings is significantly smaller than the sum of cross-sections of individual vortices. Furthermore, were two system of vortex rings have the same total line length, the Andreev scattering cross-section is significantly larger of a system consisting of bigger rings. We introduced a screening factor of a system of vortex rings and showed that it decreases strongly with the average ring's radius. Our results may be helpful for inferring quantitative properties of
ballistic vortex rings produced by a vibrating grid at its low
velocities, as in the experiment reported by Bradley {\it et al.} \cite{Bradley2}.
Our results can also be used for detecting the transition, observed
in the cited experiment, from a gas of vortex rings to the dense
vortex tangle (based on the two-dimensional model of vortex points, a qualitative
analysis of change of the Andreev reflection coefficient during such
a transition was given in our earlier paper~\cite{BSS3}; our new results may
allow a more quantitative analysis).

\acknowledgments

This work was supported by the Leverhulme Trust, grant numbers F/00 125/AH and F/00 125/AD. We are grateful
to N.~B.~Kopnin, S.~N.~Fisher, and M.~Krusius for discussions.


\begin{thebibliography}{99}
\bibitem{sizes}
Our choice of the range for vortex ring's sizes has been motivated by experiments~\cite{Bradley2} where the average size of rings produced by the vibrating grid was about $5\times10^{-4}\,{\rm cm}$. Besides, the considered range of sizes was also chosen to capture a change of behavior, due to screening effects, of the angle-averaged cross-section occuring in the case where the ring becomes sufficiently small, see Fig.~\ref{fig:aversigmas}. For rings larger than the largest considered in this work, $R\approx4\times10^{-3}\,{\rm cm}$ the angle-averaged cross-section remains linear with $R$ and is well approximated by formula~(\ref{eq:lindepend}).
\end{thebibliography}
\end{document}